\documentclass[preprint,showpacs,prb,floatfix,superscriptaddress,citeautoscript 
,cite]{revtex4}
\usepackage{graphicx}
\usepackage{color}
\usepackage{amsmath}

\usepackage{booktabs}

\begin{document}

\title{Hexagonal AlN: Dimensional-Crossover-Driven Bandgap Transition}

\author{C. Bacaksiz} 
\affiliation{Department of Physics, Izmir
Institute of Technology, 35430 Izmir, Turkey}

\author{H. Sahin}
\email{hasan.sahin@uantwerpen.be}
\affiliation{Department of Physics, University of Antwerp, 2610,
Antwerp, Belgium}

\author{H. D. Ozaydin} 
\affiliation{Department of Physics, Izmir
Institute of Technology, 35430 Izmir, Turkey}

\author{S. Horzum}
\affiliation{Department of Physics, University of Antwerp, 2610,
Antwerp, Belgium}

\author{R. T. Senger}
\affiliation{Department of Physics, Izmir Institute of Technology,
35430 Izmir, Turkey}

\author{F. M. Peeters}
\affiliation{Department of Physics, University of Antwerp, 2610
Antwerp, Belgium}

\date{\today}


\begin{abstract}

Motivated by a recent experiment that reported the successful synthesis of 
hexagonal 
(\textit{h}) AlN [Tsipas \textit{et al.} Appl. Phys. Lett. \textbf{103}, 251605 
(2013)] we investigate structural, electronic and vibrational properties of 
bulk, bilayer and monolayer structures of \textit{h}-AlN by using 
first-principles calculations. We 
show that the hexagonal phase of the bulk  $h$-AlN is a stable direct-bandgap semiconductor. Calculated 
phonon spectrum displays a rigid-layer shear mode at 274 cm$^{-1}$ and an 
E$_{g}$ mode at 703 cm$^{-1}$ which are observable by Raman measurements. In 
addition, single layer $h$-AlN is an indirect-bandgap semiconductor with a 
nonmagnetic ground state. For the bilayer structure, AA$^{\prime}$ type 
stacking is found to be the most favorable one and interlayer interaction is 
strong. While N-layered $h$-AlN is an indirect bandgap semiconductor for 
N=1-10, we predict that thicker structures (N$>$10) have a direct-bandgap at 
the $\Gamma$-point. The number-of-layer-dependent bandgap transitions in 
$h$-AlN is interesting in that it is significantly different from the 
indirect-to-direct crossover obtained in the transition metal dichalcogenides.  

\end{abstract}

\maketitle

\section{Introduction}

Bulk structures of III-V semiconductors have been widely studied due to their 
importance for technological applications such as blue light-emitting diodes, 
lasers operating in the blue and the ultraviolet regime, high temperature 
diodes and
transistors\cite{yeh1,yeh2,morkoc,vurgaftman,fukumoto,christensen}. In 
addition, technological advances lead to the emergence of novel low-dimensional 
forms of III-V binary compounds. Experimental fabrication of AlN 
nanowires\cite{wu1,duan,zhao,li}, nanobelts\cite{wu2,tang,yu} and 
nanodots\cite{goh,bouch} have already been reported. In recent years, 
theoretical and experimental studies of graphene\cite{novo1} provided a wide 
range of knowledge for a new class of materials and have opened up 
possibilities for the synthesis of many similar structures such as; silicene 
\cite{kara,liu,cahangirov,sw}, germanene \cite{cahangirov,ni,dav,yang}, 
transition metal dichalcogenides (TMDs) 
\cite{mx2-0,mx2-1,mx2-2,mx2-3,exp-ws,st-mose,hs-wse,hs-res} and hexagonal
structures of III-V binary compounds (e.g. $h$-BN, $h$-AlN)
\cite{hasan1,zhuang-hennig,wang,kim,farahani}. Among the binary compounds, 
after single layer BN, AlN is currently a material of much interest due to its 
semiconducting nature and its large bandgap which is suitable for device 
applications.

Following the first theoretical study on the stability of hexagonal AlN 
reported by Sahin \textit{et al.}\cite{hasan1}, successful experimental 
realization of hexagonal AlN phase was achieved very recently by Tsipas 
\textit{et al.}\cite{tsipas} In the early study of Du \textit{et al.}, 
energetic and electronic properties of one-dimensional AlN nanostructures such 
as nanowires with hexagonal cross sections, double-and triple-walled faceted 
nanotubes and single-walled faceted AlN nanotubes were investigated.\cite{du} 
Between the first theoretical prediction and the experimental realization of 
$h$-AlN, other groups have also focused on this material. Zheng \textit{et 
al.}\cite{zheng} predicted that zigzag AlN nanoribbons have an indirect bandgap 
whereas  armchair AlN nanoribbons have a direct bandgap, and these bandgaps 
monotonically decrease with increasing ribbon width. Almeida \textit{et 
al.}\cite{almeida} studied the energetics and electronic properties of typical 
defects in a $h$-AlN network such as vacancies, anti-sites and impurities. It 
was shown that defects such as N vacancies and Si impurities lead to the 
breaking of the planar symmetry of the $h$-AlN sheet and significant changes in 
the band structure in the vicinity of the Fermi level. In addition, Chen 
\textit{et al.} systematically investigated the electronic structure of 
armchair and zigzag AlN, GaN nanoribbons and investigated also the electronic 
properties of AlN/GaN nanoribbon heterojunctions. They found that the bandgap 
of both nanoribbons decrease monotonically as the ribbon widths increase and 
that the bandgap of the nanoribbon heterojunctions are closely related to the 
AlN/GaN ratio\cite{chen}. Shi \textit{et al.} calculated the magnetic 
properties of undoped and transition metal (TM) doped AlN nanosheets by using 
first-principles calculations. They reported that AlN nanosheet is nonmagnetic, 
whereas a single $3d$ TM atom can bring about large local magnetic moments in 
TM-doped AlN nanosheets\cite{shi}.

\begin{figure}
\includegraphics[width=12cm]{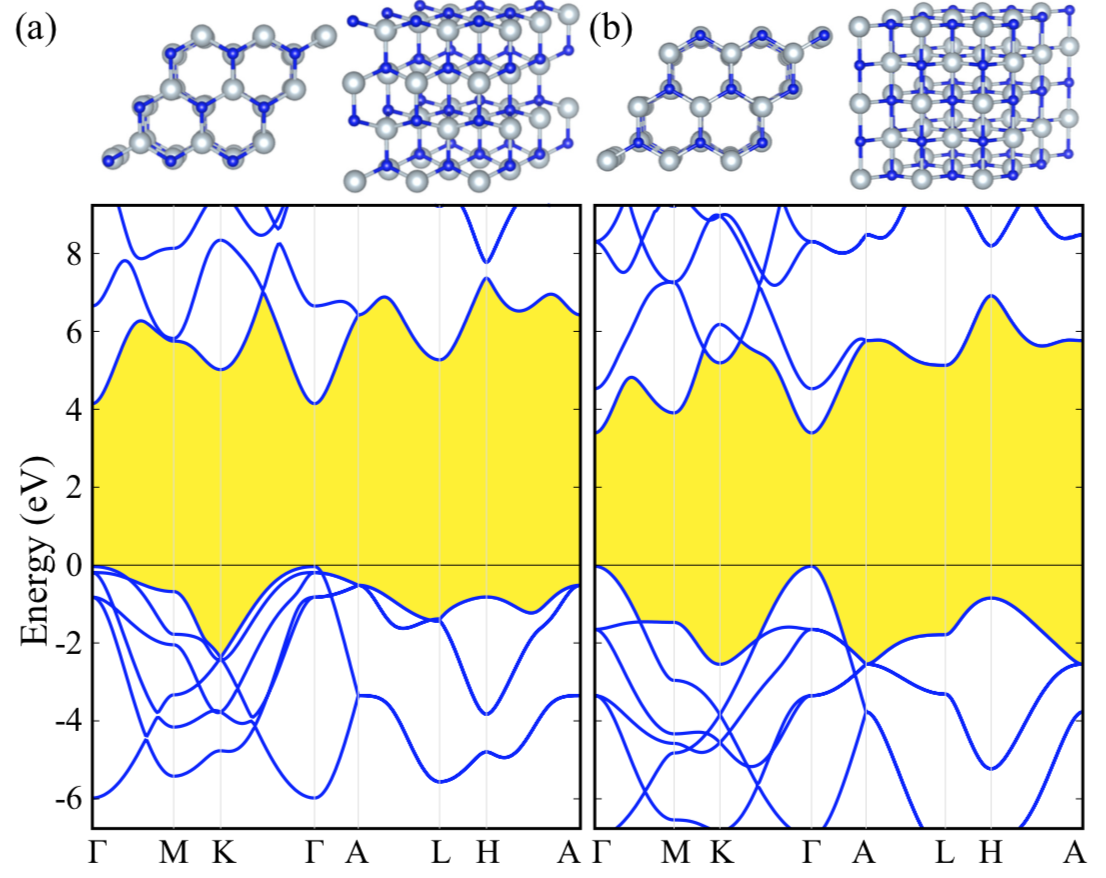}
\caption{\label{f1}
(Color online) Top and tilted-side views of atomic structures of bulk (a) 
wurtzite and (b) hexagonal AlN (top panel). 
Their electronic band dispersion (lower panel).}
\end{figure}

Motivated by the recent study of Tsipas \textit{et al.}\cite{tsipas} reporting 
the formation of stable \textit{h}-AlN phases at the early stages of AlN growth, 
we aim to address the following experimental observations: (i) modification in 
electronic valence band structure, (ii) relatively large lattice parameter that 
was reported as an indication of \textit{h} phase, (iii) reduction in band gap 
in comparison with bulk wurtzite, and (iv) dynamical stability of 
\textit{h}-AlN. 

Our study reveal that the synthesized $h$-AlN, although it has structural 
similarities, electronically displays different characteristics from $h$-BN and 
TMDs. As well as weak van der Waals (vdW) interactions, ionic character of the interlayer 
interactions plays an important role in the electronic properties of 
mutilayered $h$-AlN. Therefore, while monolayers of  $h$-BN and TMDs are direct 
band gap semiconductors, $h$-AlN has an indirect band gap. In addition, upon 
dimensional reduction from bulk to bilayer energy the band gap decreases like 
as was found for $h$-BN and TMDs.

The paper is organized as follows: in Sec. \ref{comp} we give details of our 
computational methodology. In Sec. \ref{bulk} we briefly overview the 
structural and electronic properties of hexagonal AlN (\textit{h}-AlN) 
structure together with wurtzite structure. In Sec. \ref{thin} stability of few 
layer structures and their electronic properties are investigated in details. 
Our findings are discussed in Sec. \ref{conc}

\section{Computational Methodology}\label{comp}
Our investigations of the minimum energy configurations of different AlN 
structures and their electronic properties, were carried out using the Vienna 
ab-initio simulation package VASP.\cite{vasp1,vasp2,vasp3,vasp4} which is based 
on density functional theory (DFT). The VASP code solves the Kohn-Sham 
equations\cite{Kohn-Sham} for a system with periodic boundary conditions using 
plane-wave basis set iteratively. The Perdew-Burke-Ernzerhof (PBE) form of the 
Generalized Gradient Approximation (GGA)\cite{GGA-PBE1,GGA-PBE2} was adopted to 
describe electron exchange and correlation. The band gap underestimation in 
bare GGA calculations are also examined by using the hybrid DFT-HSE06 
functional.\cite{HSE06} The vdW interaction that is 
significant for layered $h$-AlN material was taken into account for all 
multilayer structures\cite{vdW1,vdW2}. For the charge transfer analysis the 
Bader technique was used\cite{Henkelman}.

The kinetic energy cut-off of the plane-wave basis set was $500$ eV in all 
calculations. The optimization of atomic positions was performed by minimizing 
the total energy and the forces on the atoms hence the energy difference 
between sequential steps was taken $10^{-5}$ eV as a convergence criterion in 
the structural relaxation, and the convergence for the Hellmann-Feynman forces 
on each atom was taken to be $0.05$ eV/\AA{}. In addition, Gaussian smearing 
factor of $0.05$ eV was used for non-self-consistent calculations and the 
pressures on the unit cell were decreased to a value less then 1.0 kB in all 
three directions. For the determination of accurate charge densities, Brillouin 
zone integration was performed using a $35\times35\times1$ Monkhorst-Pack mesh 
for the primitive unit cell\cite{Monkhorst}. To avoid interaction between 
adjacent AlN monolayers and few layer systems, our calculations were performed 
with a large unit cell including $\sim14$ \AA{} vacuum space. We also 
calculated the cohesive energy (E$_{coh}$) which was formulated as 
$\mbox{E}_{coh} = \left( n_{Al}\mbox{E}_{Al} + 
n_{N}\mbox{E}_{N}-\mbox{E}_{T}\right) /\left( 
n_{Al}+n_{N}\right)$, where $\mbox{E}_{T}$, $\mbox{E}_{Al}$, $\mbox{E}_{N}$, 
$n_{Al}$, and $n_{N}$ are the total energy per unitcell, the energy of free Al 
atom, the energy of free N atom, number of Al and N atoms in unitcell, 
respectively.

\section{Structures of Bulk AlN}\label{bulk}

First we briefly overview the characteristics of wurtzite and recently 
synthesized $h$-AlN before the extensive investigation of the number of layer 
dependent properties of $h$-AlN are undertaken. At ambient conditions, AlN 
crystallizes in a
wurtzite ($wz$) structure that belongs to the $P6_{3}mc$ space group. As shown 
in Table \ref{table1}, PBE approximation gives the following structural 
parameters $a=3.11$~ \AA{} and $\frac{c}{a}=1.61$. In the $wz$ structure of AlN 
each Al-N bond is formed by 2.35 $e^{-}$ 
charge donation from Al to N atom and therefore it has a highly ionic 
character. In parallel with the abundance 
of the $wz$ form of AlN in nature, among the possible bulk structures $wz$ is 
the energetically most favorable one. The electronic dispersion shown in Fig. 
\ref{f1} indicates that $wz$-AlN is a semiconductor with a $4.2$ eV direct 
bandgap at the $\Gamma$ point.

%%%%%%%%%%%%%%%%%%%%%%%%%%%%%%%%%%%%%%%%%%%%%%%%%%%%%%%%%%%%%%%%%%%%%%%%%%%%%%%%
\begin{table}[htbp]
\caption{\label{table1} Calculated lattice parameter in the lateral direction 
$a$ 
and lattice parameter in the vertical direction $c$, the distance between 
layers 
d$_{LL}$, the intralayer atomic distance between Al and N atoms d$_{Al-N}$, 
the charge transfer from Al atom to N atom $\Delta\rho$, the cohesive energy 
E$_{coh}$ and the energy bandgap of the structure E$_{gap}$. (d) and (i) 
indicate direct and indirect bandgap, respectively.}
\begin{tabular}{rcccccccc}
\hline\hline
 & a   & c & d$_{LL}$ & d$_{Al-N}$ & $\Delta\rho$ & E$_{coh}$ & E$_{gap}$ 
\\
 & (\AA) & (\AA) & (\AA) & (\AA) & ($e$) & (eV) & (eV) \\
\hline
Bulk $wz$-AlN & 3.11 & 5.01 &  -   & 1.90  & 2.35 & 6.14 & 4.2 (d) \\
Bulk $h$-AlN  & 3.30 & 4.15 & 2.08 & 1.90  & 2.37 & 6.02 & 3.4 (d) \\
2L $h$-AlN    & 3.20 &   -  & 2.13 & 1.85  & 2.32 & 5.73 & 3.5 (i) \\
1L $h$-AlN    & 3.13 &   -  &  -   & 1.81  & 2.28 & 5.36 & 2.9 (i) \\
\hline\hline
\end{tabular}
\end{table}
%%%%%%%%%%%%%%%%%%%%%%%%%%%%%%%%%%%%%%%%%%%%%%%%%%%%%%%%%%%%%%%%%%%%%%%%%%%%%%%%

 In addition to some early theoretical 
predictions\cite{hasan1} the stability of the hexagonal phase of AlN ($h$-AlN) 
was proven by a very recent experimental study.\cite{tsipas} Regarding the less 
layered structures or surfaces of wurtzite materials, transformation from 
wurtzite to a graphite-like structure that allows the removal of destabilizing 
dipoles is energetically more favorable.\cite{freeman} This graphite-like 
hexagonal structure of AlN belongs to the $P6_{3}/mmc$ space group. The layered 
planar structure has two atoms in the unitcell which has lattice vectors 
$\textbf{v}_{1}=a(\frac{1}{2},\frac{\sqrt{3}}{2},0)$,
$\textbf{v}_{2}=a(\frac{1}{2},-\frac{\sqrt{3}}{2},0)$ and
$\textbf{v}_{3}=c(0,0,1)$ where $|\textbf{v}_{1}|=|\textbf{v}_{2}|$. Atomic 
coordinates are $(\frac{|v_{1}|}{3},\frac{|v_{1}|}{3},0)$ and 
$(\frac{2|v_{1}|}{3},\frac{2|v_{1}|}{3},0)$, for first and second type of atoms 
respectively. We considered that the layers are AA$^{\prime}$ stacking 
(deciding the stacking of layers is explained in Section IV-B) which is shown 
in Fig.\ref{f1}(b). And we found that the lattice constant is $a=3.30$ \AA{} 
and the distance between layers is $c=2.08$ \AA{} which is smaller when 
compared to graphite and $h$-BN ($\sim3.33$ \AA{} and $\sim3.30$ \AA{} for 
graphite\cite{Baskin,Chung} and $h$-BN\cite{Paszkowicz,Marini,Shi}, 
respectively). Intraleyer atomic distance between Al and N atoms is $1.90$ 
\AA{}. This is the same as for $wz$-AlN. The cohesive energies 
are given in Table \ref{table1}.

As shown in Fig. \ref{f1}(d), layered $h$-AlN is a direct bandgap 
semiconductor which has $3.4$ eV bandgap with band edges located at the 
$\Gamma$ point. Therefore, our results confirm the experimentally reported 
bandgap reduction\cite{tsipas} in $h$-AlN compared to bulk $wz$-AlN. At the 
top of the VB, there is a single parabolic band for holes and the main 
contributions are from the $p_{z}$ orbital of N atoms. The bottom of the CB, 
there is a single parabolic band and most of the states come from $p_{x}$ 
orbital of N atoms. Final charges of Al and N atoms are 0.63 $e^{-}$ and 7.37 
$e^{-}$, respectively. Therefore, 2.37 $e^{-}$ charges are transferred  from Al 
to N atom. Following section is devoted to the understanding of the 
characteristics of monolayer $h$-AlN and the nature of the inter-layer 
interactions.

\begin{figure}[htbp]
\includegraphics[width=11 cm]{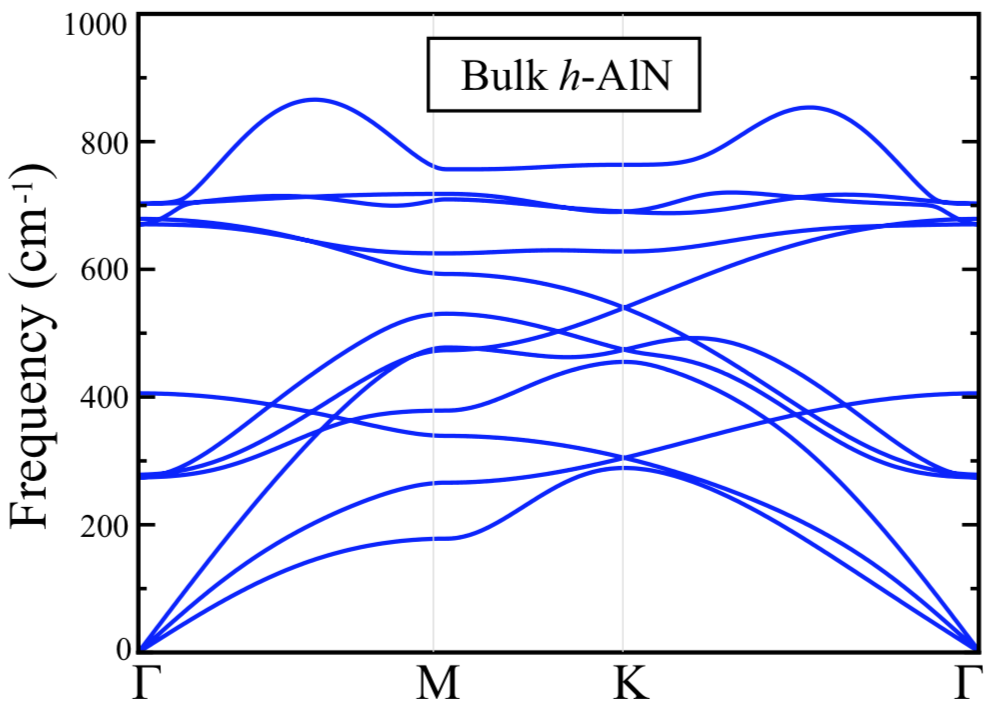}
\caption{\label{phonon}
(Color online) Phonon dispersion of bulk $h$-AlN.}
\end{figure}

\begin{figure}
\includegraphics[width=11 cm]{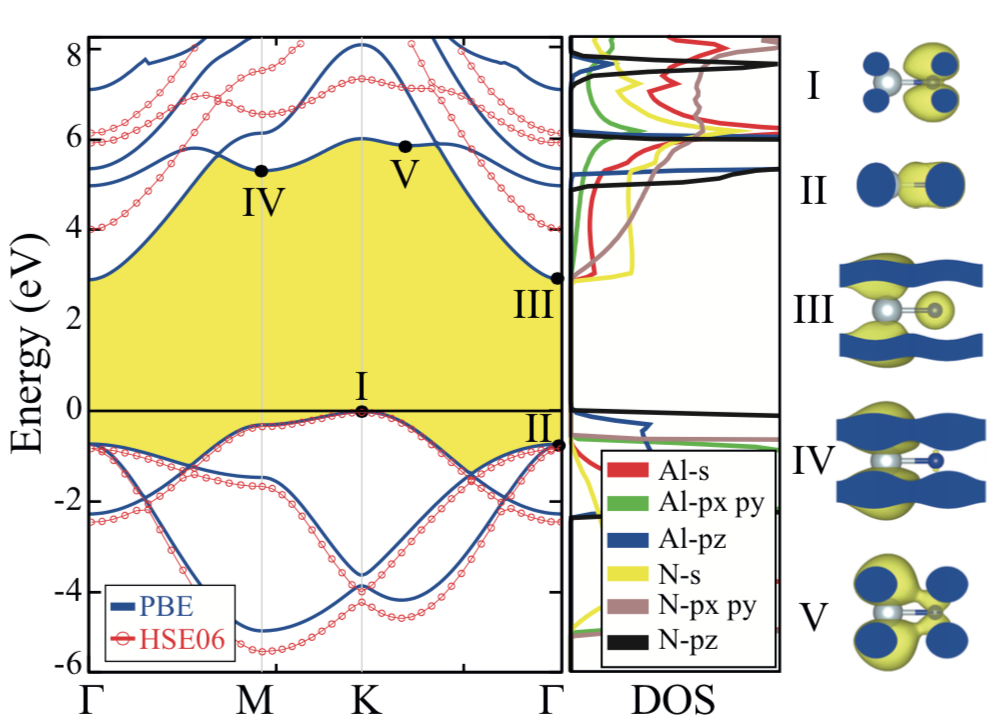}
\caption{\label{1layer}
(Color online) Monolayer $h$-AlN band diagram within PBE  and 
HSE06  (left panel), partial density of states (middle panel) and the 
band decomposed charge densities of the band edges (right panel). Sections of 
3D charge density that represent the connection to the neighboring cells are 
filled with blue color.}
\end{figure}

Lastly, we analyze vibrational properties of the $h$-AlN phase by 
calculating the phonon spectrum in whole Brillouin 
Zone (BZ) using the small displacement methodology\cite{alfe} for 32-atomic 
supercell. As shown in Fig. \ref{phonon} phonon spectrum of $h$-AlN  displays 
real eigenfrequencies in the whole BZ. Therefore, there is 
no doubt 
on the dynamical stability of the $h$-AlN structure. Here we also see that 
differing from its $h$-BN counterpart the optical and acoustical modes 
couple with each other at certain points in the BZ. In the case of $h$-AlN, 
there 
are 4 atoms per primitive unit-cell and 12 phonon branches. The 
first three are acoustical phonon modes, the phonon branch at 274 cm$^{-1}$ is 
a doubly degenerate rigid layer shear mode, known as low-energy E$_{g}$ mode, 
involving the out-of-phase motion of atoms in adjacent planes. In addition, the 
highly dispersive phonon modes with eigenvalues 406 and 679 cm$^{-1}$ 
at the zone center correspond to out-of-phase motion of Al and N atoms in 
adjacent layers. Interestingly, the contribution of the second atom type in 
these modes are negligibly small. The highest optical mode, doubly degenerate 
at the $\Gamma$ point, have the E$_{g}$ symmetry and therefore it is expected 
to be measured in Raman experiments.

\section{Monolayer, Bilayer and Few Layer $h$-A$\textbf{l}$N}\label{thin}

\subsection{Monolayer $h$-AlN}

The first prediction of the dynamical stability and electronic properties of 
single layers of $h$-AlN and similar III-V compounds were first reported by 
Sahin \textit{et al.}\cite{hasan1} Monolayer hexagonal structure 
belongs to the space group $P6_{3}/mmc$ with unit vectors 
$\textbf{v}_{1}=a(\frac{1}{2},\frac{\sqrt{3}}{2},0)$,
$\textbf{v}_{2}=a(\frac{1}{2},-\frac{\sqrt{3}}{2},0)$ where
$|\textbf{v}_{1}|=|\textbf{v}_{2}|$. In this configuration the atomic 
coordinates are given as $(\frac{|v_{1}|}{3},\frac{|v_{1}|}{3},0)$ and 
$(\frac{2|v_{1}|}{3},\frac{2|v_{1}|}{3},0)$ for first and second type of atoms, 
respectively. We calculated that the lattice constant of  monolayer 
$h$-AlN is $3.13$ \AA{} and the distance between Al and N atoms is $1.81$ 
\AA{} which is the lowest value when compared with bulk forms and bilayer 
structure 
as seen from  Table \ref{table1}. These results are in good agreement with 
the measured lattice parameters (3.13 \AA) of epitaxially grown of $h$-AlN, at 
early stages, by Tsipas \textit{et al.}\cite{tsipas}. The cohesive energy of 
monolayer $h$-AlN is the lowest among the possible phases. It is also seen that 
the amount of charge transfer from Al to N slightly decreases from bulk to 
monolayer $h$-AlN.

\begin{figure}
\includegraphics[width=12 cm]{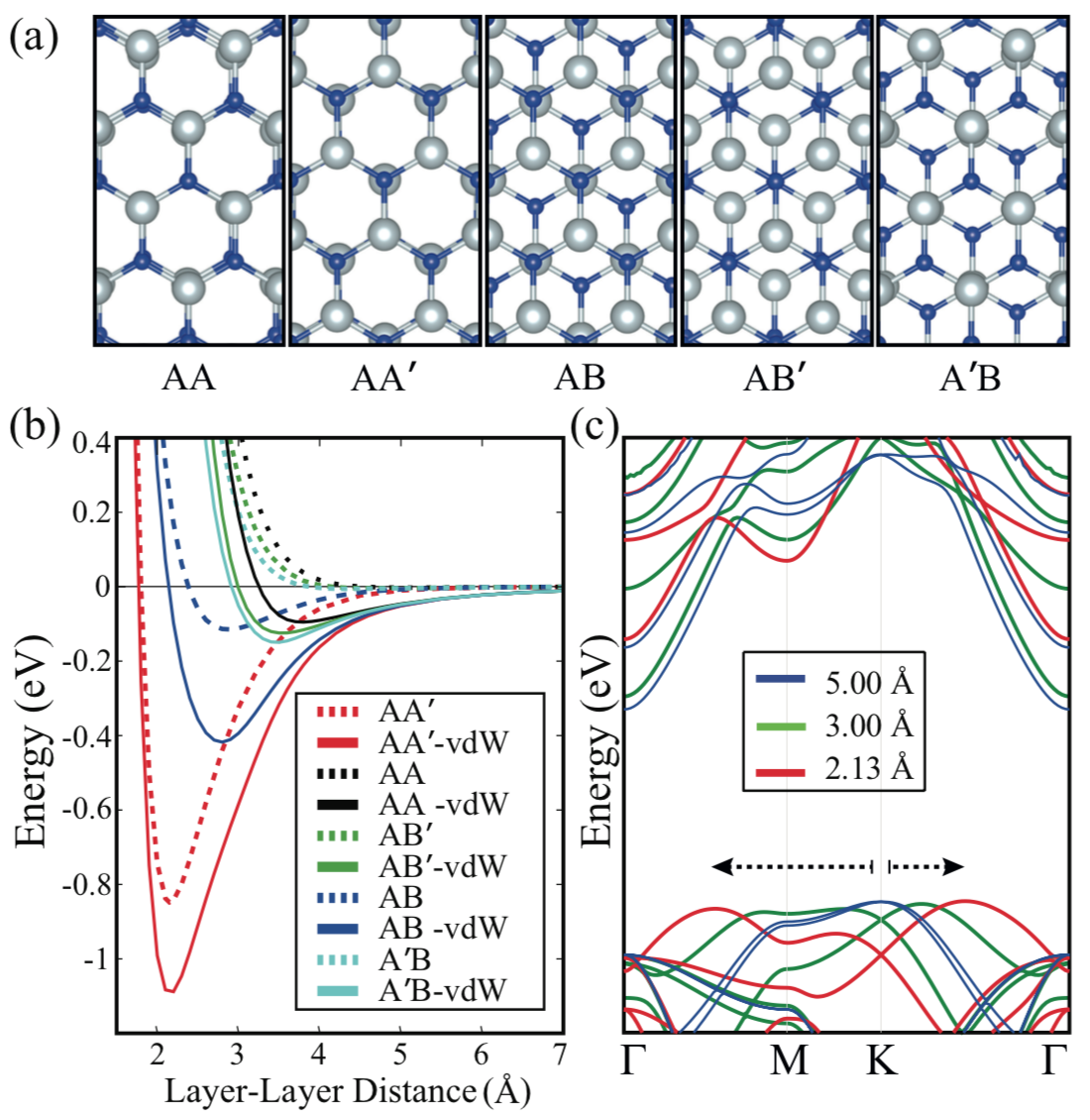}
\caption{\label{stac}
(Color online) (a) Possible stackings between two layers of $h$-AlN. (b) 
Layer-layer interaction energy for different stackings. (c) Interlayer spacing 
dependent band dispersion of AA$^{\prime}$ stacked two layers of $h$-AlN.}
\end{figure}

Fig. \ref{1layer} illustrates the band diagrams, density of states 
and also the 3D charge densities of states at the band edges for monolayer $h$-AlN. 
Differing from bulk $h$-AlN which has a direct bandgap at the $\Gamma$, the VBM 
of the monolayer is at the K point and the CBM is at the $\Gamma$ point, 
therefore monolayer $h$-AlN is an indirect bandgap insulator with $2.9$ eV 
bandgap. It is also seen that bare-GGA and DFT-HSE06 
approximated electronic band dispersions shown in Fig. \ref{1layer} are almost 
the same. With DFT-HSE06 hybrid functional, $h$-AlN is an indirect bandgap 
material with a bandgap of $4.06$ eV. It is also noteworthy to mention that 
$h$-BN is the counterpart of AlN with a direct bandgap at the K point. 

Right panel of Fig.\ref{1layer} shows that the bonding states at VBM (label I) 
are mainly composed of N-$p_{z}$ orbitals and the degenerate state, with lower 
energy, at the $\Gamma$ point (label II) stems from hybridized $p_{x}$ and 
$p_{y}$ orbitals. In addition, the CBM located at the $\Gamma$ point has a 
quadratic-like dispersion (label III) and the relevant charge density is in the 
form of surface state. Therefore, the conduction electrons in single layer 
$h$-AlN display free-electron-like behavior. Moreover the anti-bonding 
states with higher energy at the M and the K points are composed by mainly Al 
and slightly N states together.

\begin{figure}
\includegraphics[width=10 cm]{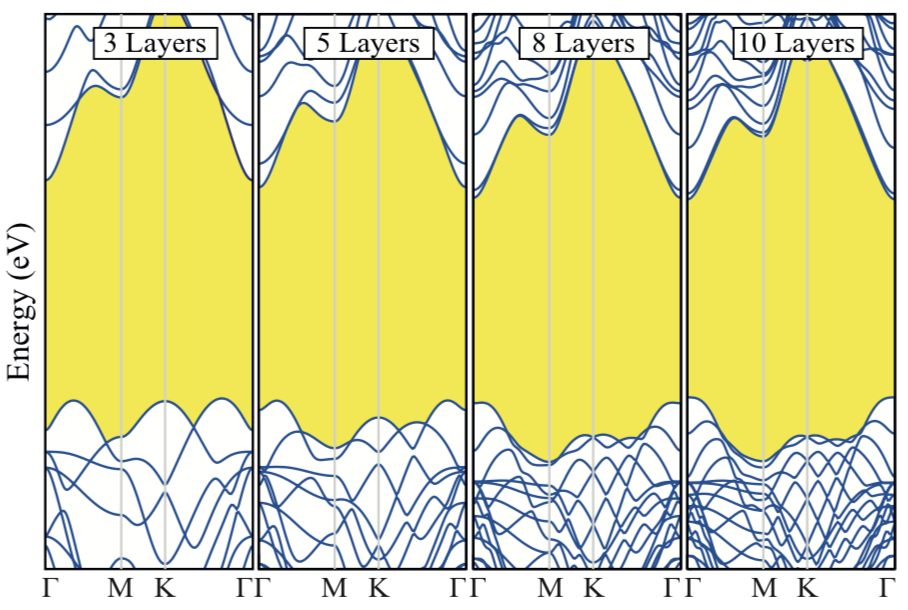}
\caption{\label{multilayer}
(Color online) Evolution of electronic band dispersion of N-layer $h$-AlN.}
\end{figure}

\subsection{Bilayer $h$-AlN and Layer-layer Interaction}

During the growth  of lamellar materials prediction of the most 
preferable stacking sequence is of importance. To determine the stacking order 
of bilayer AlN, we examined five possible (see Fig. \ref{stac}(a)) stacking 
types: AA (Al over Al and N over N), AA$^{\prime}$ (Al over N and N over Al), 
AB (Al over N and N over center of hexagon), A$^{\prime}$B (Al over Al and N 
over center of hexagon), AB$^{\prime}$ (N over N and Al over center of 
hexagon). 

To determine the interaction strength and interlayer interaction profile of 
above-mentioned stacking orders of bilayer $h$-AlN, total energies were 
calculated as a function of layer-layer distance. As seen from Fig. 
\ref{stac}(b), the largest interlayer coupling energy and shortest interlayer 
distance is obtained for AA$^{\prime}$ type stacking. Therefore AA$^{\prime}$ 
stacking corresponds to the ground state stacking order. It is also seen that 
the vdW correction to the interaction energy is $\sim$250 meV 
per unitcell. In addition, for AA$^{\prime}$ stacking, layer-layer distance is 
$2.13$ \AA{} which is the minimum value among all stacking combinations. Here 
it appears that due to the presence of surface states interlayer distance 
in bilayer $h$-AlN is slightly larger than that of bulk $h$-AlN. However, 
compared to graphite\cite{Baskin,Chung} and $h$-BN\cite{Paszkowicz,Marini,Shi}, 
and their bilayer forms, $h$-AlN has a smaller interlayer distance. The lattice 
parameter, $3.20$ \AA{}, which is the 
maximum value of all stacking combinations that slightly differs from the value 
obtained by self-correlation analysis of the experimental data (3.14$\pm$0.06) 
for few layers. This slight disagreement can be understood from the effect of 
the Ag(111) substrate.

\begin{table}
\caption{\label{table2} Calculated direct and indirect transition energies 
between the
VB and CB edges of single to ten layered structures.}
\begin{tabular}{rcccccccc}
\hline\hline
           & $\Gamma \longrightarrow \Gamma$ & 
K$\longrightarrow \Gamma$ & 
$\Gamma$M$\longrightarrow\Gamma$ & 
K$\Gamma\longrightarrow\Gamma$ & E$_{coh}$ & & & \\
\hline
1L AlN     & 3.62 & 2.92 & -    & -    &5.36 & & &\\
2L AlN     & 4.29 & -    & 3.61 & 3.53 &5.73 & & &\\
3L AlN     & 4.05 & 3.58 & 3.57 & 3.53 &5.82 & & &\\
4L AlN     & 3.79 & -    & 3.51 & 3.50 &5.86 & & &\\
5L AlN     & 3.61 & 3.73 & 3.46 & 3.46 &5.89 & & &\\
6L AlN     & 3.48 & -    & 3.41 & 3.41 &5.92 & & &\\
7L AlN     & 3.38 & 3.77 & 3.35 & 3.35 &5.93 & & &\\
8L AlN     & 3.33 & -    & 3.30 & 3.30 &5.94 & & &\\
9L AlN     & 3.27 & 3.81 & 3.26 & 3.26 &5.95 & & &\\
10L AlN    & 3.22 & -    & -    & -    &5.96 & & &\\
\hline\hline
\end{tabular}
\end{table}

Moreover, in Fig. \ref{stac}(c), we analyze how the electronic band dispersion 
of AA$^{\prime}$-stacked bilayer is modified with varying interlayer distance. 
Negligibly interacting, weakly interacting and highest interaction (ground 
state) cases are illustrated using layer-layer distances of 5.00, 3.00 and 
2.13 \AA{}, respectively. The band diagram of negligibly interacting case 
resembles the electronic structure of monolayer $h$-AlN. However, when the 
layers start to interact, for instance in the weakly interacting case 
(delineated by green lines in Fig. \ref{stac}(c)), the top of the VBM are 
shifted towards the $\Gamma$ zone center. When the interlayer distance reaches 
 its optimum position, one of the VBM appears at K-$\Gamma$ while the 
other one is located at M-$\Gamma$. Interestingly, the location of CBM in BZ 
is independent from the interlayer distance. Hence from monolayer 
to bilayer, the indirect character is conserved and the bandgap is changed 
from 2.9 to $3.5$ eV.

\subsection{Few Layer $h$-AlN}
\begin{figure}
\includegraphics[width=10 cm]{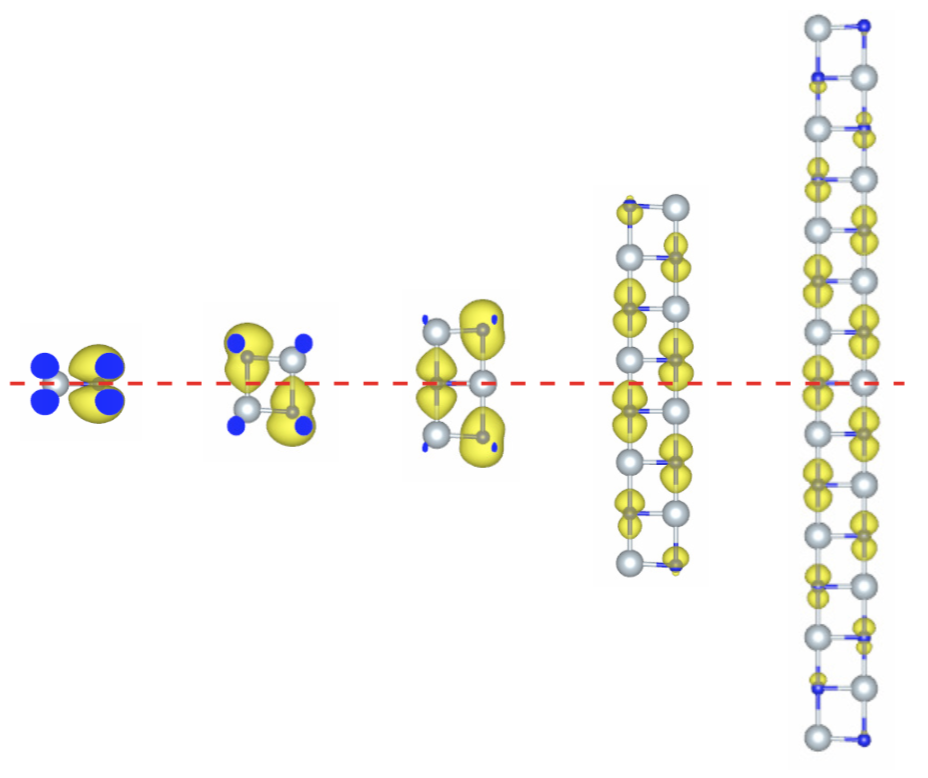}
\caption{\label{multilayer2}
(Color online) The evolution of the bonding charge density of $h$-AlN for 
monolayer, bilayer, 3-layer, 8-layer and 15-layer. Sections of 3D charge 
density that represent the connection to the neighboring cells are filled with 
blue color.}
\end{figure}

Synthesis of stable ultra-thin $h$-AlN structures, sub-monolayer to 12 layers, 
was demonstrated by Tsipas \textit{et al.}\cite{tsipas}, recently. Our 
first-principles total energy optimization calculations revealed that lamellar 
hexagonal structure of N-layered (N=1,2,...) AlN is a stable phase. As shown in 
Figs. \ref{multilayer} and \ref{multilayer2}, planarity of the few-layered 
structures is preserved, except the negligible buckling at the surface layers, 
while electronically few layer $h$-AlN is significantly different from 
monolayer one. 

In order to illustrate how the electronic structure evolves with increasing 
number of layers we present the band dispersion of 3, 5, 8 and 10-layered 
$h$-AlN in Fig. \ref{multilayer}. Here there are several interesting trends that 
is worth to note: (i) the energy of the conduction band edge at the $K$ point 
decreases with increasing number of layers, (ii) states at conduction band edge 
which have surface state character remain unchanged. The energy difference 
between the band edges VBM($K$) and CBM($\Gamma$) increases from 2.92 eV for 
monolayer 
to 3.81 eV for 10-layer $h$-AlN, and (iii) the most significant influence of 
additional number of layers is on the bonding states forming the valence band 
maximum at the $\Gamma$M and the $\Gamma$K points. It appears that due to the 
strong interlayer interaction each additional layer shifts these band edges 
towards the zone center. This gradual shift of VB edges continues up to 10 
layers. Upon the formation of the 10th layer $h$-AlN structure reaches the bulk 
limit and it becomes a direct band gap semiconductor. Increasing the thickness 
furthermore has no influence on the electronic properties of N-layered $h$-AlN 
and they are direct bandgap ($3.22$ eV for ten layered) materials like 
bulk $h$-AlN. As given in detail in Table \ref{table2}, transition energies 
between the points converge asymptotically to values of bulk $h$-AlN. Increasing 
the number of layers, cohesive energies converge to that of bulk $h$-AlN. It 
appears that only 
few layer $h$-AlN structures with thickness N$>$11 are suitable for lasing 
device applications. Here calculated modifications in electronic structure 
explains and supports the experimental findings of Tsipas \textit{et 
al.}\cite{tsipas}.

However, the evolution of the VBM with changing number of layers requires 
further attention. In Fig. \ref{multilayer2} we present the 3D charge density 
of the electronic states located at the VBM. It is clearly seen that the 
character of 
the bonding state (VBM) is modified monotonically with increasing number 
of layers. With increasing number of layers the hybridization between 
N-$p_{z}$ states is weakened. In addition, due to the mild buckling the 
interaction of the surface states which have $sp^{3}$-like character and 
N-$p_{z}$ 
states vanishes at the vicinity of the upper- and the lower-most layers. 
Therefore for 
enough thick $h$-AlN materials, which is found to be larger than 10-layers, the
electronic structure is mainly determined by uniformly distributed N-$p_{z}$ 
states and hence the structure behaves like a bulk material having a direct 
bandgap at the $\Gamma$ symmetry point.

\section{Conclusions}\label{conc}

Using first principles methodology, we investigated the thickness dependent 
electronic properties of layered hexagonal AlN. First, we showed that the 
bulk structure of hexagonal AlN is a semiconductor with a direct 
bandgap at the $\Gamma$ point. Phonon 
spectrum analysis of this structure reveals two Raman-active modes at 274 and 
703 cm$^{-1}$. Here, the lattice parameter of $h$-AlN, which is larger than the 
wurtzite phase, agrees with the experiment of Tsipas \textit{et
al.}\cite{tsipas} that reports the formation of hexagonal regions at the early 
stages of the growth process. Next, the formation of multilayer structures and 
most favorable stacking order were investigated via total energy calculations. 
It is seen that, similar to the hexagonal BN counterpart, among the possible 
stackings AA$^{\prime}$ type stacking is the most favorable. It is worthwhile 
to note that the interlayer interaction is larger compared to similar layered 
materials with hexagonal lattice symmetry. Differing from some other 
two-dimensional crystal structures such as graphite, $h$-BN and TMDs, here the 
ionic interlayer interaction is dominant between the $h$-AlN layers. 
Therefore, for the synthesis of $h$-AlN the epitaxial growth technique 
performed by Tsipas \textit{et al.}\cite{tsipas} appears more suitable than 
mechanical exfoliation.

Subsequently, the evolution of the electronic structure of N-layered $h$-AlN 
was investigated for structures with 1-15 layers. It is seen that unlike 
bulk $h$-AlN, monolayer  $h$-AlN has an indirect bandgap where VBM and CBM are 
at the K and the $\Gamma$ points, respectively. Moreover, it is seen that upon 
the formation of additional layers  valence band edges gradually shift towards 
the Brillouin Zone center. Such a modification of the valence band states is in 
good agreement with the experiment of Tsipas \textit{et al.}\cite{tsipas} It 
was also calculated that the few layered structures reach the bulk limit and 
their electronic properties remains almost unchanged after the formation of 10 
layers. Therefore, structures thicker than 10 layers, exhibits a direct band gap 
of 3.22 eV at the $\Gamma$ point. We believe that easy synthesis and the 
presence of a thickness-dependent bandgap crossover in few-layered hexagonal 
AlN structures are very important for novel device applications.

\begin{acknowledgments}

This work was supported by the Flemish Science Foundation (FWO-Vl) and
the Methusalem foundation of the Flemish government. Computational
resources were provided by TUBITAK ULAKBIM, High Performance and Grid
Computing Center (TR-Grid e-Infrastructure). C.B. and R.T.S. acknowledge the 
support from TUBITAK Project No 114F397. H.S. is supported by a
FWO Pegasus Long Marie Curie Fellowship.

\end{acknowledgments}

\end{document}